\documentclass[sigconf]{acmart}
\AtBeginDocument{%
  }

\copyrightyear{2026}
\acmYear{2026}
\setcopyright{cc}
\setcctype{by}
\acmConference[UIST '26]{The 39th Annual ACM Symposium on User Interface Software and Technology}{November 02--05, 2026}{Detroit, MI, USA}
\acmBooktitle{The 39th Annual ACM Symposium on User Interface Software and Technology (UIST '26), November 02--05, 2026, Detroit, MI, USA}
\acmDOI{10.1145/3830398.3830597}
\acmISBN{979-8-4007-2856-3/2026/11}

\usepackage{multirow}
\usepackage{siunitx}
\usepackage{subcaption}

\newif\ifcomment
\commentfalse

\newif\ifrevision
\revisiontrue

\ifcomment

\ifrevision

\newcommand{\added}[1]{\textcolor[rgb]{0, 0.094, 0.89}{#1}}
\newcommand{\deleted}[1]{}

\else

\newcommand{\added}[1]{\textcolor[rgb]{0.1, 0.56, 1}{#1}}
\newcommand{\deleted}[1]{\textcolor[rgb]{0.7,0.,0.}{{#1}}}

\fi

\else
\newcommand{\added}[1]{#1}
\newcommand{\deleted}[1]{}
\fi

\begin{document}

\title{FleetScape: A Mixed Reality Sandtable for Spatial Supervision and Control of Scalable Drone Fleets}

\author{Peisen Xu}
\email{ps.xu@nus.edu.sg}
\orcid{0000-0003-1312-3061}
\affiliation{%
\institution{National University of Singapore, IPAL}
\institution{CNRS@CREATE}
  \country{Singapore}
}

\author{Jérémie Garcia}
\email{jeremie.garcia@enac.fr}
\orcid{0000-0001-7076-6229}
\affiliation{%
  \institution{University of Toulouse, Fédération ENAC ISAE-SUPAERO ONERA}
  \city{Toulouse}
  \country{France}
}

\author{Peter Cleveland}
\email{petercleveland@nus.edu.sg}
\orcid{0009-0004-8343-242X}
\affiliation{%
\institution{CNRS, IPAL, Singapore}
  \institution{National University of Singapore, Augmented Human Lab}
  \country{Singapore}
}

\author{Wei Tsang Ooi}
\email{ooiwt@comp.nus.edu.sg}
\orcid{0000-0001-8994-1736}
\affiliation{%
  \institution{National University of Singapore, School of Computing}
  \country{Singapore}
}

\author{Christophe Jouffrais}
\email{christophe.jouffrais@cnrs.fr}
\orcid{0000-0002-0768-1019}
\affiliation{%
  \institution{CNRS, IPAL}
  \country{Singapore}
}
\affiliation{%
  \institution{Univ of Toulouse, IRIT}
  \city{Toulouse}
  \country{France}
}

\renewcommand{\shortauthors}{Xu et al.}

\begin{teaserfigure}
  \includegraphics[width=\textwidth]{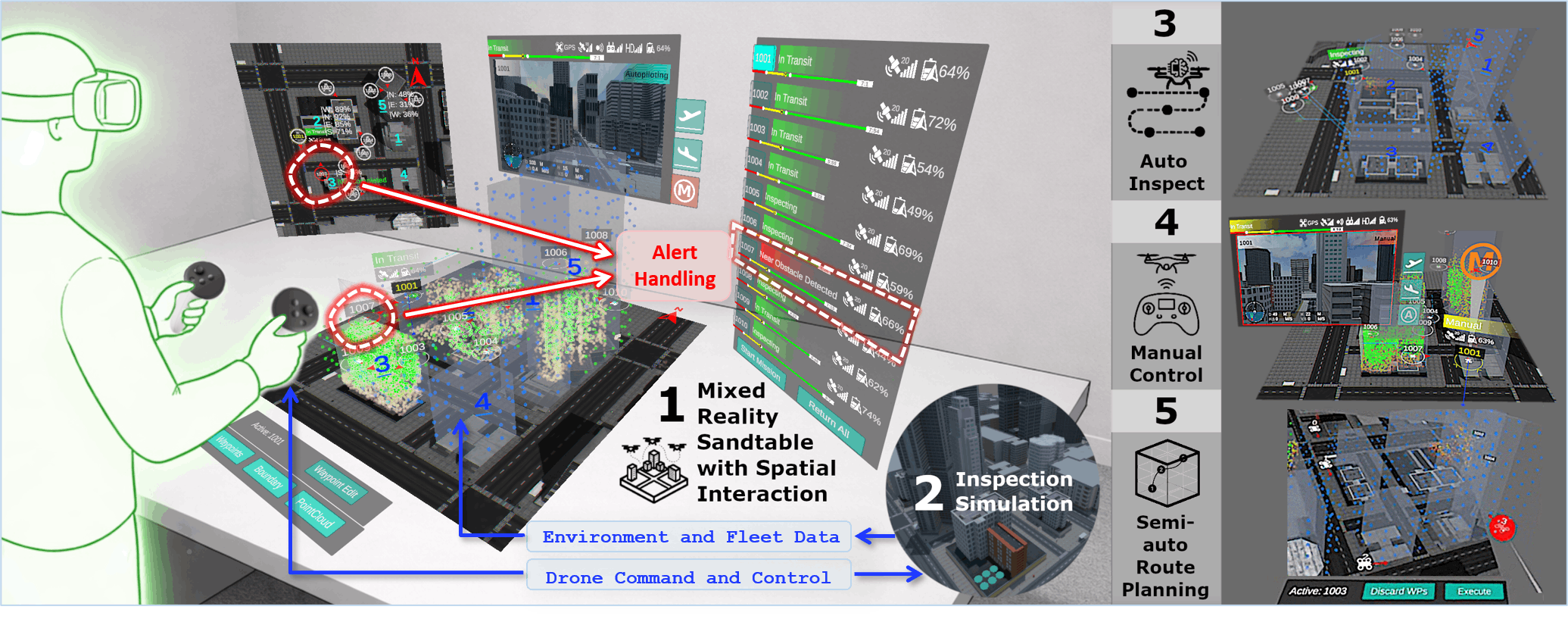}
  \caption{Overview of FleetScape. (1) A MR interface featuring a 3D Sandtable that enables spatial interaction and supports situational awareness by mapping fleet states, alerts, and mission data to their precise locations. (2) An Inspection Simulation providing a real-time environment (using urban assets from City Pack \textcopyright{Mindravel Interactive}~\cite{mindravel-city-pack-modular-and-tileable}) and fleet data while receiving drone commands. FleetScape supports three control modes: (3) Autonomous Inspection; (4) Direct Manual Control via camera feed; and (5) Semi-autonomous Route Planning for interactive waypoint editing. }
  \footnotetext{Urban 3D assets: ``City Pack -- Modular and Tileable'' (Mindravel Interactive), Unity Asset Store, used under the Unity Asset Store EULA.}
  \Description{A diagram showing a person wearing a mixed reality headset using the FleetScape interface. On the left, the operator stands before a 3D Sandtable with a vertical Fleet Status Board on the right and a 2D Minimap on the left, labeled component 1 (Mixed Reality Sandtable with Spatial Interaction). In the center, a box labeled 2 (Inspection Simulation) receives drone command and control from the interface and returns environment and fleet data via bidirectional arrows. On the right, three smaller screenshots illustrate the three control modalities: 3 (Auto Inspect) showing autonomous drone flight paths on the sandtable, 4 (Manual Control) showing a drone camera feed alongside the Drone Control Panel, and 5 (Semi-auto Route Planning) showing custom waypoint editing within the sandtable.}
  \label{fig:teaser}
\end{teaserfigure}
\begin{abstract}

As autonomous drone deployments scale from individual units to coordinated swarms, the human operator’s role shifts from direct piloting to high-level supervision. Current interfaces often treat multi-drone control as a scaled-up version of single-drone operation. We instead investigate how reframing fleet supervision as spatial interaction can better support the spatial, temporal, and safety demands of complex missions. We present FleetScape, a Mixed Reality (MR) sandtable system that externalizes layered real-time mission, safety, and environmental data while enabling fluid transitions between manual intervention and autonomous supervision. We developed a high-fidelity building inspection simulation that generates and streams synchronized multi-drone and environmental data for MR visualizations. We used this prototype to conduct a user study with six experienced drone pilots managing fleets of up to 15 drones. Our findings show that FleetScape supports situational awareness through layered spatial representations and clarifies control mode transitions. However, a limit to situational awareness was observed as fleet size increases, leading to different supervisory strategies. Finally, we derive design implications for supporting scalable drone fleet supervision. 

\end{abstract}




\maketitle

\section{Introduction}

There is a growing social need for drone sensing abilities in domains such as building inspection and digital twin generation~\cite{srivastava2022uav, salem2024strategic, aais2022buzzing}. Automating drone fleets can significantly improve the efficiency of these tasks~\cite{cao2025cooperative}. However, most multi-drone or swarm systems currently focus on functionality rather than usability~\cite{hocraffer2017meta}, while real-world drone operations are often dynamic due to environmental conditions, regulations, and specific mission requirements~\cite{rakotonarivo2023cleared}.
These challenges motivate the need for human-in-the-loop approaches that allow operators to adapt system behavior to dynamic conditions.

Combining automated planning and human-guided systems has shown improvements in both the safety and effectiveness of swarm operations~\cite{kira2009exerting,cummings2011impact}. However, scaling from single-drone to multiple-drone operation substantially increases cognitive load due to multitasking demands~\cite{hocraffer2017meta,hoang2023}. Piloting a single drone through a traditional first-person camera view is already cognitively demanding~\cite{kocer2022immersive}, making it difficult to scale to multiple drones~\cite{menda2011optical}. In highly automated systems, operators shift from direct control to supervision, introducing new cognitive and mission management challenges~\cite{cummings2007automation, hocraffer2017meta}. 

3D spatial interfaces have been shown to improve single-drone teleoperation compared to conventional 2D interfaces, but few studies explore their application for coordinated multi-drone supervision~\cite{kolling2015human, hocraffer2017meta}. Specifically, it is unclear how spatial interfaces can present concurrent fleet and mission information while facilitating seamless transition between global supervision and direct control with map manipulation~\cite{roldan2017multirobot, walker2024immersive}. This limits the understanding of operators' supervisory strategies~\cite{endsley2017here}. This gap motivates our central research question: \textit{How can we reconfigure drone fleet supervision as spatial interaction to improve situational awareness and support the transitions between direct control and supervision?}

In this paper, we introduce FleetScape, a Mixed Reality (MR) interface centered on a 3D Sandtable, an exocentric spatial miniature of the operational environment, to support a single operator supervising a drone fleet \added{in a stable and structured built environment}, see Figure~\ref{fig:teaser}. Unlike prior immersive VR swarm interfaces~\cite{roldan2017multirobot, walker2024immersive}, FleetScape enables seamless switching between supervising the entire fleet and directly controlling an individual drone by adjusting the sandtable’s scale and focus, as well as smooth transitions between manual piloting, route planning, and autopilot modes. Our system also integrates multiple spatial data layers (Building Boundaries, Inspection Waypoints, Point Clouds, and Drone Miniatures) complemented by a 2D Minimap, Status board, and Control Panel.

To evaluate FleetScape, we built a multi-drone simulator generating real-time 3D data streams and safety events for customizable and realistic building inspection scenarios\added{, which assume careful mission planning with safety procedure}. We conducted an exploratory user study with six professional pilots to assess mission and safety situational awareness, using the Situation Awareness Global Assessment Technique (SAGAT)~\cite{endsley1988situation}, across increasing fleet sizes. Results indicate that pilots can effectively control and supervise 5-10 drones while maintaining situational awareness. 

Beyond this range, cognitive load increases and operators shift toward reactive, exception-based monitoring, highlighting key design challenges for scalable fleet control interfaces.


This work makes the following contributions: (i) A spatial interaction design space supporting mission and safety situational awareness for fleet supervision and individual drone control; (ii) FleetScape, a 3D mixed reality sandtable interface enabling layered spatial visualization and transitions across control modes; and (iii) an evaluation with six professional pilots 
identifying limits to manageable fleet size, revealing spatial supervision strategies, and informing the design of scalable drone fleet supervision interfaces.


\section{Related Work}

\subsection{Situational Awareness and Interfaces for Multi-Drone Supervision and Control}
Situational awareness (SA) refers to the operator’s ability to perceive relevant elements in the environment (level 1), understand their meaning (level 2), and anticipate their future state (level 3)~\cite{endlsey1995b}. Prior research in human–robot interaction identifies situational awareness and workload management as central challenges when supervising multiple autonomous systems~\cite{kolling2015human, hocraffer2017meta, chen2014human}. As the size of the fleet increases, operators must alternate attention between individual drones and higher-level mission objectives, which can increase cognitive workload and degrade situational awareness~\cite{agrawal2020}. However, many existing interfaces present this information through fragmented views such as separate camera feeds, telemetry panels, and 2D maps, making it difficult to maintain a coherent spatial understanding of the fleet and mission context~\cite{chen2014human, agrawal2020, hoang2023}. This also leads to mental rotation problems, where operators must mentally align a view with a representation stored in memory and constantly translate their desired movement into a control input that matches the drone's current orientation~\cite{shepard1971mental, chen2021pinpointfly, cho2017fly}. To quantify scalability, researchers use the ``Fan-out'' metric based on Activity Time (AT) and Interaction Time (IT) to estimate the maximum number of drones a single person can control~\cite{cummings2008predicting, aubuchon2022multi}. 

\subsection{Immersive Interface for Drone Control}

Immersive technologies, such as Virtual Reality (VR) and Mixed Reality (MR), have been increasingly adopted to address the spatial limitations of 2D interfaces~\cite{green2008human}. For single drone control, MR interfaces have demonstrated significant benefits in reducing cognitive load and improving spatial awareness by embedding navigational telemetry and camera feeds directly into the user's field of view~\cite{kocer2022immersive, xu2025safespect}. In multi-robot contexts, VR interfaces allow operators to immerse themselves in a digital twin of the environment, facilitating more intuitive spatial interactions and mission planning~\cite{roldan2017multirobot}. Despite these advantages, fully immersive VR can isolate operators from their physical surroundings. Although MR provides a hybrid approach that preserves real-world context, the majority of MR drone interfaces remain centered on direct manual piloting instead of higher-level supervisory control over entire fleets~\cite{tan2025intelligent, xu2025safespect,victor2025iguana}. 

\subsection{Spatial Interaction for Robot Supervision}

The Spatial Interaction paradigm draws on an operator’s natural spatial reasoning to control digital content, employing multimodal inputs to manipulate volumetric displays, position widgets, perform 3D selections, or add annotations~\cite{besanccon2021state}. In robot supervision, spatial interaction is often reflected in a sandtable metaphor, which functions as a tangible or digital, high‑resolution medium that closes gaps of distance and viewpoint for high-level tactical planning and geographic situational awareness~\cite{piper2002illuminating, kolling2015human}. According to domain experts, the sandtable substantially reduces the "guesswork" about elevation and terrain patterns relative to traditional 2D maps~\cite{li2015flying}.  
Digital applications like Augmented Reality Sandtables (ARES) and Immersive Interaction Interface (I3) have been developed explicitly for commanding units and conducting military-style mission planning on defined spatial volumes~\cite{amburn2015augmented,walker2024immersive}. 
These systems demonstrate the value of spatially grounded interaction for planning and coordination.
However, existing sandtable interfaces often focus on ground-based robots or abstract swarm behaviors, lacking the multi-layered vertical spatial data (such as 3D trajectories, dynamic obstacles, and building fa\c{c}ades) and transition between various levels of control required for supervising coordinated drone fleets in complex aerial environments.

\subsection{Control Transitions in Spatial Interfaces}

Human-automation collaboration is often classified by a level-of-autonomy (LOA) taxonomy spanning information acquisition, analysis, decision selection, and action implementation. Existing spatial interfaces largely anchor operators at fixed points on this continuum. SafeSpect~\cite{xu2025safespect} and IGUANA~\cite{victor2025iguana} center on direct manual control of a single vehicle, while Flying Frustum~\cite{li2015flying}, and Tan et al.~\cite{tan2025intelligent}, focus on semi-autonomous path planning. ARES~\cite{amburn2015augmented} and I3~\cite{walker2024immersive} support fleet-level mission planning but do not offer per-vehicle control without abandoning the mission context. 

Control transitions are among the most cognitively demanding events in supervisory control~\cite{grindley2024handover}, prone to two key failures: the \textit{out-of-the-loop} problem~\cite{endsley1995ootl}, where reduced engagement impairs timely intervention~\cite{endsley2017here, agrawal2020}, and \textit{mode error}~\cite{sarter1995mode}, where the system behaves contrary to operator expectations. However, existing systems rarely support fluid transitions across levels of autonomy within a unified spatial context to encourage engagement, while maintaining situational awareness and minimizing mode confusion.


\section{FleetScape: Design and Implementation}
To design FleetScape, we followed an iterative, expert informed process grounded in prior work on multi-drone supervision~\cite{xu2025safespect,agrawal2020,hoang2023}. We conducted a two hour design workshop with two professional drone pilots using an early 3D sandtable mock-up to explore the limitations of existing interfaces for multi-drone control and how a spatial interface could help. The workshop revealed key challenges in representing spatial relationships and transitioning between levels of control with existing interfaces. From these insights, we derived a design space that guided the development of FleetScape.

\subsection{Design Space}

We structured FleetScape around three design dimensions (Table~\ref{tab:design_space}) that support drone fleet supervision as spatial interaction.

\textbf{Interface modalities} specify how map and fleet status are presented and managed across coordinated modalities, from a Sandtable (3D) for spatial reasoning, a Minimap (2D) for stable overviews, and a Fleet Status Board (1D) for compact textual status. Prior studies show the effectiveness of combining 3D and 2D displays for estimating position, orientation, and volume~\cite{tory2005visualization}. Our design also aligns with guidelines for using multiple views in information visualization, including diversity (through the use of multiple levels of abstraction in each view), complementarity (consistent icons and color-coding across views), and decomposition (offloading complex status data from Sandtable to Fleet Status Board)~\cite{wang2000guidelines}. 


\textbf{Spatial visualizations} focus on how mission, drone, and environmental information are projected on the 3D Sandtable. This information includes Building Boundaries, Inspection Waypoints, Point Clouds, and Drone Miniatures that are co-located within a unified spatial representation. This allows operators to perceive relationships between drones, tasks, and constraints. 

\textbf{Control transitions} support seamless switching between levels of automation, from Manual Control to Semi-autonomous Route Planning and Auto Inspection. This is enabled through consistent mode visualizations, continuous visual shifts from global to localized spatial contexts, focused control of a single drone, and on-the-fly re-routing integrated in mission flow. 

\begin{table*}[t]
  \caption{Overview of the FleetScape Design Space}
  \label{tab:design_space}
  \begin{tabular}{@{}p{2cm}p{3.5cm}p{11cm}@{}}
    \toprule
    \textbf{Dimension} & \textbf{Category} & \textbf{Focus / Components} \\
    \midrule
    \textbf{1. Interface} 
    & Sandtable (3D) & Situated 3D context for real-time spatial awareness \\
    \textbf{Modalities} 
    & Minimap (2D) & Stable top-down fleet overview (drone positions, mission progress)\\
    & Fleet Status Board (1D) & Text-based fleet-wide telemetry and alerts \\
    \midrule
    \textbf{2. Spatial \newline Visualizations} 
    & Building Boundary & Preplanned 3D boundaries that convey safety constraints and surface completion status. Color encodings distinguish unfinished and completed surfaces. \\
    & Inspection Waypoints & Preplanned waypoints traversed by the autonomous inspection algorithm, indicating visited, unvisited, and assigned states. \\
    & Point Clouds & Real-time point clouds generated from building scans; color indicates point confidence. \\
    & Drone Miniatures & 3D drone miniature showing position, heading, view frustum, collision ring, and warning status. The controller-focused drone is differentiated by visualization granularity. \\
    \midrule
    \textbf{3. Control \newline Transitions} 
    & Manual Control & Dual-stick control supported by a Drone Control Panel with camera feed and mode-switching buttons. \\
    & Auto Inspect & Full autonomous flight. States visualized via colored trajectories on the Sandtable, icons on the Minimap, and text on the Fleet Status Board.\\
    & Semi-auto Route Planning & Custom waypoint-edit mode for adjusting drone flight paths. \\
    \bottomrule
  \end{tabular}
\end{table*}

\subsection{Mission Workflow}

FleetScape supports a typical inspection workflow consisting of four stages: (i) Before the mission, the operator configures the building boundaries with inspection waypoints and loads them in the system. (ii) The operator initiates the operation from the Fleet Status Board, and the drones start autonomous inspection. (iii) The operator supervises the fleet on the Sandtable with supporting views, selecting drones, monitoring progress, and intervening when needed through manual control or waypoint adjustments. The system continuously updates mission progress, drone status, and environmental data, triggering alerts for critical events. The drones automatically return to their homepoints when required (e.g., low battery). (iv) The mission ends when all target surfaces are inspected or all drones have returned home.
The operator can also terminate the mission at any time by recalling all drones.



\subsection{System Overview}


FleetScape runs as a standalone mixed reality application on a Meta Quest 3 headset and receives drone telemetry and environment data via Wi-Fi. \added{Prior work uses free-hand interaction~\cite{tan2025intelligent, victor2025iguana} and motion controllers~\cite{walker2024immersive, xu2025safespect}. Free-hand interaction is natural for map manipulation and waypoint placement, but can be imprecise for manual flight~\cite{victor2025iguana}. For this reason, we prioritized input with two Meta Touch Pro controllers.}

\subsection{Interaction Modes}

FleetScape supports two primary interaction modes:  \textbf{Bound interaction mode}, which helps the user maintain situational awareness by customized placement of different views in the peripheral vision. Users can reposition and scale interface elements in space using grab interactions and joysticks. This is consistent with the interaction paradigm of Meta Horizon OS. \textbf{Map interaction mode} encourages spatial exploration of the maps for fleet supervision. Users directly manipulate spatial representations (e.g., panning, zooming, rotating the Sandtable or Minimap). These interactions enable users to flexibly align viewpoints and focus on relevant parts of the mission space. \deleted{The user can toggle modes using the primary button of the controller. }

\added{During the iterative design process, map manipulation was changed from a two-handed trigger and grip gesture, to a mode toggle with the controller's primary button plus single-handed grip and joystick interaction, since pilots rarely moved the widget once it had been placed. This revised scheme allowed both hands to act in parallel and better supported transitions between map interaction and drone control.}

\begin{figure*}[t]
  \centering
  \Description{Interface components showing six different views: Fleet Status Board, 2D Minimap, Drone Control Panel, and three different 3D Sandtable visualizations showing inspection waypoints, building boundary, and point clouds.}
  \includegraphics[width=\textwidth]{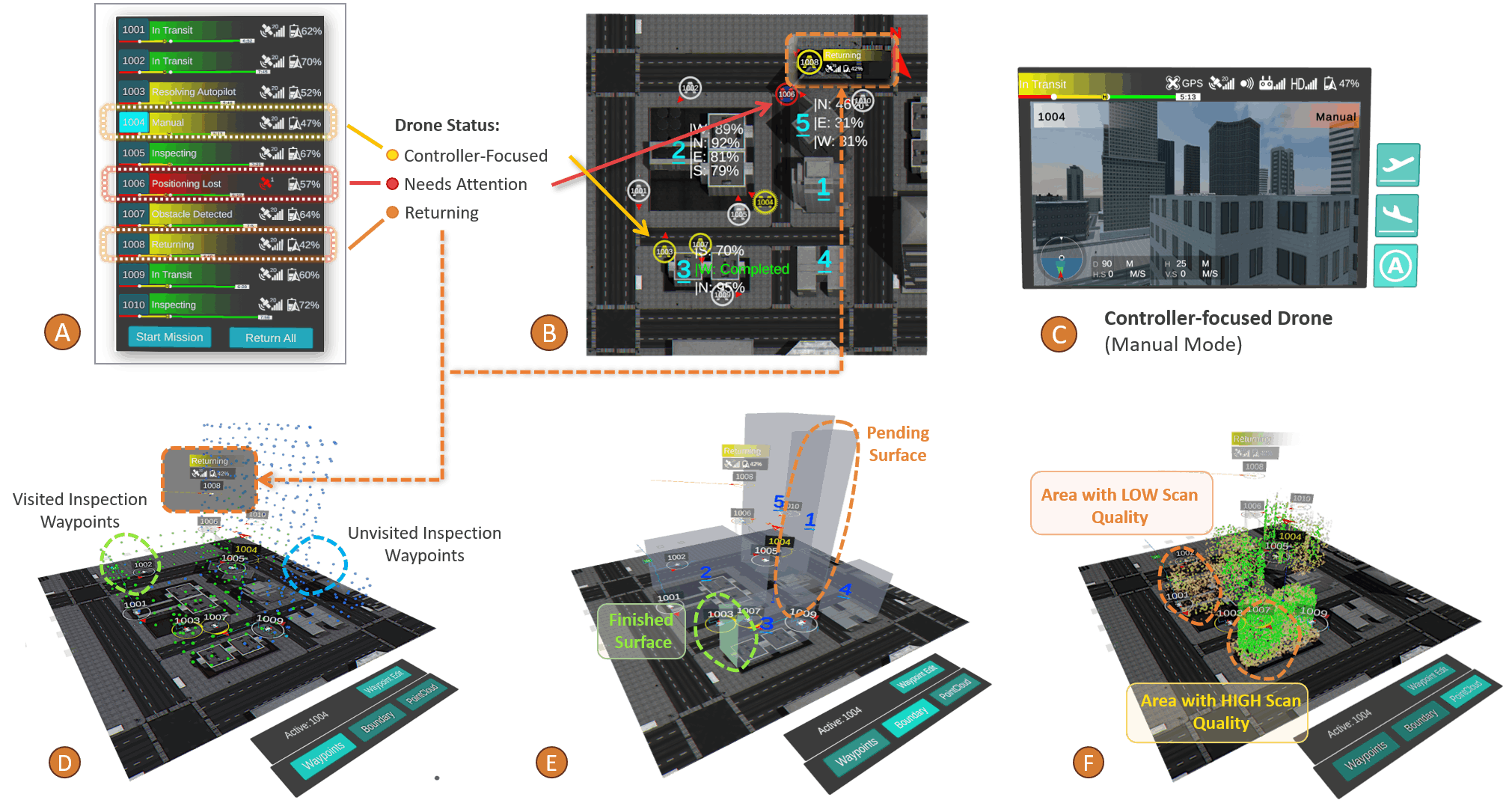}
  \caption{Interface components of FleetScape. A. Fleet Status Board; B. Minimap; C. Drone Control Panel. Sandtable with D. Inspection Waypoints; E. Building Boundary; F. Point Clouds. The simulated environment is based on city pack from Unity Asset Store \textcopyright{Mindravel Interactive}}
  \label{fig:design_details}
\end{figure*}

\subsection{Interface Components}

\subsubsection{Sandtable}

The Sandtable is the central component of FleetScape. It represents a spatial miniature of the operational environment, including drones, mission progress, and environmental constraints, within a unified 3D space. This allows operators to directly perceive and reason about fleet behavior and spatial structure.

Multiple layers of information are integrated into the Sandtable, including Building Boundaries, Inspection Waypoints with drone trajectories, real-time Point Clouds, and Drone Miniatures. 
Visual encodings (e.g., color and level of detail) communicate task progress, assignment, and system state as illustrated in Figure~\ref{fig:design_details}.D/E/F.
Point Cloud data is streamed and rendered in real time within the Sandtable to support environmental awareness. To enable efficient visualization of large and dynamic datasets, we use a GPU-based pipeline that updates point positions and confidence values with compute shaders and visualizes using a visual effect graph. Points are visually encoded by confidence, allowing operators to assess scan quality at a glance. During interaction (e.g., scaling or rotating the Sandtable), visualization is clipped to the Sandtable boundary. 

Drone Miniatures follow a focus-aware design to support scalable supervision. The selected (controller-focused) drone is rendered with detailed information, including orientation, view frustum, and status, following a safety-first design adapted from~\cite{xu2025safespect}, while non-focused drones are abstracted to simplified visual markers (e.g., rings and projections) to reduce clutter. 

Drone Miniatures outside the current focus region are hidden. Visual encodings such as color and indicators reflect drone state and alert levels, supporting rapid identification of critical events. When entering manual control, the Sandtable automatically centers on the selected drone. In addition, operators can also dynamically adjust the scale and orientation of the Sandtable, enabling a seamless transition from global supervision to local interaction.

\subsubsection{Supporting Views}

\added{We first explored using the Sandtable as the sole central workspace. However, pilot feedback showed that operators also needed stable fleet-wide summaries, rapid attention switching, and compact alert or status information. Thus, }FleetScape complements the Sandtable with two additional views. The Minimap (Figure~\ref{fig:design_details}.B) provides a stable top-down overview of fleet distribution and mission progress, while the Fleet Status Board (Figure~\ref{fig:design_details}.A) presents compact textual information on drone states and alerts. Notifications are displayed as a head-up overlay in the corner of the view. Audio cues are provided for key events, including drone warnings, surface completion, return-to-home initiation, and drone landing. These views and audio alerts support quick reference and decision-making alongside spatial interaction.
The Drone Control Panel (Figure~\ref{fig:design_details}.C) enables per-drone interaction, including mode switching, recalling to homepoint, manual takeoff, and controlling via camera view. Once a drone is selected, the Drone Control Panel streams the camera view from the controller-focused drone and presents the current status and control mode.





\subsubsection{Control Transitions}


The fluid transition across multiple levels of autonomy is enabled by four design features:

Firstly, the control state is delineated consistently across different interface components. For instance, manual intervention instantly highlights the target drone with a yellow status indicator on both the Sandtable and Minimap, while supporting textual status update on the Fleet Status Board and the Drone Control Panel to reflect the "Manual" state.

Secondly, the operator can select to focus on a specific drone via spatial interaction on the Sandtable, Minimap, or Fleet Status Board. \added{Earlier prototypes automatically centered, rotated, zoomed, and followed the selected drone. While this accelerated transitions to focused control, pilots reported that it disrupted awareness of the broader mission context. The final design therefore uses a gentler zoom-and-center transition, combined with a visual indicator above the selected drone. This preserves the benefits of rapid focus, supporting a visually continuous shift from global multi-drone supervision to highly localized spatial interaction, while reducing the risk of losing fleet-level awareness.} \deleted{When an operator selects a drone requiring intervention from the global views, the Sandtable automatically scales and centers around the target, supporting a visually continuous shift from global multi-drone supervision to highly localized spatial interaction.} 

Thirdly, recognizing the high cognitive demand of manual piloting via a first-person camera view~\cite{kocer2022immersive}, FleetScape limits manual intervention to a single, controller-focused drone. While the operator actively pilots this focal drone, the system ensures fleet continuity; non-focused drones continue operating autonomously, resolving route conflicts, returning to home, or holding position in response to critical issues, such as GPS loss. 

Finally, the system allows on the fly semi-autonomous route planning within the shared spatial context through direct manipulation. Operators can contextually re-route a drone to a new surface by placing and adjusting custom waypoints within the Sandtable. Upon execution, the drone integrates the custom waypoint sequence into its pathfinding and collision avoidance framework. Progress is visualized only when the drone is selected. Once the custom sequence is completed, the drone resumes the inspection to the nearest surface, pulling the operator's attention back to the fleet-wide supervision.




\subsection{Multiple Drone Building Inspection Simulation Environment}

To support controlled evaluation, we developed a multi-drone building inspection simulator that generates realistic mission scenarios \added{centered on routine and goal-oriented operations with safety procedures. The simulator uses building bounding boxes and an occupancy map as inputs, both of which can be derived from real world sources such as Google Maps, BIM models or site plans during the planning stage of the operations. }\deleted{streams real-time drone, mission, and environmental data, see Figure~\ref{fig:teaser}. The simulator models autonomous inspection behavior, multi-drone coordination, and urban environmental dynamics.}

The simulator integrates modular urban assets \textcopyright{Mindravel Interactive}~\cite{mindravel-city-pack-modular-and-tileable}, drone dynamics adapted from UnityDroneSim~\cite{uavs-berkeley-unitydronesim}, sensor modules adapted from SafeSpect~\cite{xu2025safespect}, and autonomous path-finding using A-star search algorithm, together with Unity-based physical interactions. Inspired by~\cite{cao2025cooperative}, we developed a custom algorithm for mission planning and multi-drone coordination (detailed in Appendix~\ref{sec:autonomy}).

\deleted{The simulator also provides configurable parameters such as fleet size, building layout, navigation uncertainty, and battery conditions, allowing systematic variation of operational complexity. This enables controlled experimentation while preserving realism in drone behavior and mission structure.} \added{The simulator provides configurable parameters such as fleet size, building layout, and battery conditions. It also enables controlled manipulations such as inducing navigation uncertainty and simulation pauses. The navigation uncertainty is operationalized through GPS localization error and wind drift. Wind drift was modeled as randomized mechanical drift, minimal under reliable GPS conditions and amplified under GPS loss, reflecting how stabilization compensates when GPS is available. Localization was intentionally made imprecise by adding a dynamic GPS-dependent random offset and sending this sensed position value to the MR frontend. During GPS-loss events, pilots used sandtable miniatures only for rough positioning and relied on the camera view for precision.}

\section{Exploratory Evaluation with Drone Pilots}

We conducted an in-person exploratory evaluation with experienced and professional drone pilots to investigate how spatial interaction supports the supervision of multi-drone fleets, to assess the effect of fleet size on performance with our interface, and to identify potential areas to improve the scalability of 3D spatial interactions. The study was approved by our Institutional Review Board. All participants provided informed consent prior to participation.

\subsection{Method}

\subsubsection{Stages and Participants}

We tested our system in two stages. In the first stage (trial stage), we focused on improving the usability of the interface and validating the experiment procedures with three experienced drone pilots (one licensed pilot, two amateur pilots). In the second stage (final stage), we conducted the main user study with six expert drone pilots, all of whom had previous experience in managing complex airspace conditions. Five of them currently hold a pilot license in Singapore, demonstrating their ability to perform (1) pre-flight and post-flight checks on drones under 25 kg, (2) the technique to manually control drones smoothly under GPS-denied conditions, and (3) competence in carrying out emergency procedures~\cite{sp2025uapl}. The non-licensed pilot, included in the final stage user study, has eight years of extensive usage of drones for photography and videography, with experience in operating drones in various environments, including urban areas. Details of the participants are summarized in Table~\ref{tab:participants}.

\begin{table}[!htp]\centering
\caption{Participant demographics. Participant IDs use two suffixes: t or f denotes the study stage (trial or final), and a or l denotes pilot proficiency (amateur or licensed).}\label{tab:participants}
\footnotesize
\begin{tabular}{lcccccc}\toprule
ID & Gen. & Exp. & UAPL & Age & Ver. & Fleet Order \\\midrule
t1\_a & M & 1+ & No & 25--34 & Trial & 5-10-15 \\
t2\_a & F & 2+ & No & 25--34 & Trial & 5-10-15 \\
t3\_l & M & 5+ & Yes & 25--34 & Trial & 5-10-15 \\
f1\_l & M & 2+ & Yes & 45--54 & Final & 10-15-5 \\
f2\_l & M & 5+ & Yes & 25--34 & Final & 15-5-10 \\
f3\_l & M & 5+ & Yes & 45--54 & Final & 5-10-15 \\
f4\_l & M & 1+ & Yes & 45--54 & Final & 10-15-5 \\
f5\_l & F & 1+ & Yes & 35--44 & Final & 15-5-10 \\
f6\_a & M & 5+ & No & 25--34 & Final & 5-10-15 \\
\bottomrule
\end{tabular}
\end{table}

\subsubsection{Task}

The simulated environment in which all participants conducted flight operations was designed to replicate a typical city center. It featured buildings positioned close to each other, varying in size, and including several with complex geometries. In the trial stage, we thoroughly tested the pilots' supervision limits with different operational complexities. We finalized the region of interest as 5 buildings with a total of 18 surfaces and 532 inspection waypoints. A theoretical optimal fleet size was computed based on the fan-out theory~\cite{olsen2004fan}, detailed in Appendix~\ref{sec:fleet-size-determination}. Participants completed the final evaluation with three different fleet sizes consisting of 5, 10, and 15 drones.


Participants were assigned two primary objectives. First, they were responsible for ensuring fleet safety by continuously monitoring the status of each drone and responding promptly to issues such as GPS errors or collision warnings. Second, they were tasked with maximizing mission throughput by keeping the drones active and covering as many inspection waypoints as possible within a 7-minute time limit. Participants performed several drone interactions to achieve these goals: launching drones after charging, manually piloting drones when necessary, and rerouting drones to improve inspection coverage across building surfaces. 


\subsubsection{Data Collection}

We used a combination of quantitative and qualitative methods to gather comprehensive data on pilot performance and user experience, including the Situation Awareness Global Assessment Technique (SAGAT)~\cite{endsley1988situation}, performance tracking, subjective questionnaire, and semi-structured interview. 

\paragraph{Situational Awareness Measurement Freezes}

The simulation was paused by the facilitator at three predetermined stop points during each test session. During these freezes, participants answered two sets of questions targeting mission progress and drone status. These questions were designed to assess three levels of situational awareness (SA): perception, comprehension, and projection. The questions are summarized in Table~\ref{tab:sagat-questions}. The three pause timings were chosen to capture different operational contexts: (1) normal autonomous flight, (2) an active GPS error event, and (3) a GPS error coinciding with drones returning to the landing pad. Chances of GPS errors scale linearly as fleet size increases.
The SAGAT score was calculated based on the method described in Appendix~\ref{sec:sagat-scoring}.

\begin{table*}[ht]
\centering
\caption{SAGAT questions for evaluating situational awareness in fleet management. Questions are organized by two domains (Mission Progress and Drone Status) across three SA levels (Perception, Comprehension, and Projection).}
\label{tab:sagat-questions}
\begin{tabular}{p{1.8cm} p{2cm} p{11cm}}
\toprule
\textbf{Domain} & \textbf{SA Level} & \textbf{Question} \\
\midrule
\multirow{3}{*}{\parbox{1.8cm}{Mission\\Progress}}
 & Perception & Which building(s) have active drone(s) working on it? \\
\cmidrule(l){2-3}
 & Comprehension & How many buildings have finished surface(s)? \\
\cmidrule(l){2-3}
 & Projection & If operations continue unchanged, which building will be finished next? \\
\midrule
\multirow{6}{*}{\parbox{1.8cm}{Drone\\Status}}
 & \multirow{2}{*}{Perception} & How many drone(s) is in one of the stoppage statuses (GPS loss, near collision, returned and landed, manual interruption) that causes automation stoppage? \\
\cmidrule(l){2-3}
 & Comprehension & What is/are the cause(s) of automation stoppage for each of these drones? \\
\cmidrule(l){2-3}
 & \multirow{2}{*}{Projection} & If the operations continue unchanged, how many drone(s) are likely to stop the automation within the next 1 minute? \\
\bottomrule
\end{tabular}
\end{table*}

\paragraph{Subjective questionnaires}

To capture participants' subjective experiences, we utilized three standardized questionnaires following the experiment. The Situation Awareness Rating Technique (SART)~\cite{taylor2017situational} was employed to assess pilots' perceived situation awareness. SART is grounded in the theory that SA is a function of the demand on attentional resources, the supply of those resources, and the user's understanding of the situation. We also administered the Bedford Workload Scale~\cite{roscoe1990subjective}, a uni-dimensional 10-point rating scale originally developed for flight assessments, which uses a hierarchical decision tree to evaluate spare mental capacity and operator workload. Finally, the Technology Acceptance Model (TAM)~\cite{davis1989technology} was used to assess participants' perceived usefulness (PU), perceived ease of use (PEOU), attitude towards using (ATT), and their behavioral intention (BI) of the interface. The detailed questionnaires can be found in the Appendix~\ref{sec:data-collection}. 

\subsubsection{Environment and Procedure}

The user study was conducted in a quiet meeting room, providing a controlled environment with only the facilitator and participant present. It followed a structured workflow consisting of two practice sessions for familiarization with the interface, followed by three test sessions (one per fleet size condition)
, and concluded with a qualitative feedback session. 

The first practice session introduced participants to the interface and basic controls with three drones. After starting the autonomous mission, participants were asked to perform two tasks: (1) manually navigate a drone to a nearby surface, and (2) use the custom waypoint function to re-route a drone to another surface. The second practice session was designed to familiarize participants with the SAGAT procedure using five drones, during which they experienced three freeze points and answered SAGAT questions.

\begin{figure}[t]
\centering
\includegraphics[width=\columnwidth]{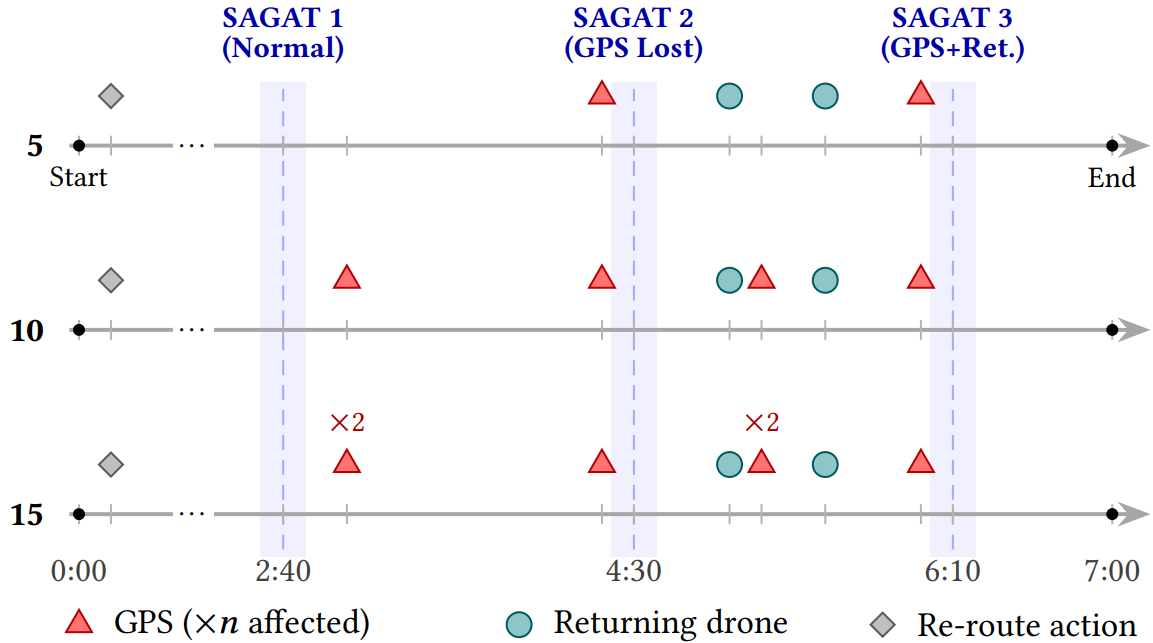}
\Description{A timeline diagram with three parallel horizontal rows labeled 5, 10, and 15 drones, spanning from mission start to the 7-minute mark. Three vertical dashed lines mark SAGAT freeze points at 2:40, 4:30, and 6:10, labeled SAGAT 1 (Normal), SAGAT 2 (GPS Lost), and SAGAT 3 (GPS + Return), respectively. Red upward-pointing triangles denote GPS disruption events; their count increases with fleet size (2 for 5 drones, 4 for 10, and 6 for 15), with some labeled ×2 to indicate two simultaneous disruptions. Teal circles mark drones returning home at 5:00 and 5:30. Gray diamonds mark pilot-initiated re-route actions, which typically occur near the beginning of the mission.}
\caption{Test session timeline for the three fleet size conditions. Vertical dashed lines indicate SAGAT freeze points. GPS disruptions (red triangles) vary according to fleet size. Teal circles denote approximate drone return timings. Re-route actions (gray diamonds) are typically initiated by pilots at the beginning of the mission.}
\label{fig:test-session-timeline}
\end{figure}

During the test sessions, participants completed the mission under one of three fleet size conditions (5, 10, or 15 drones), with the order of conditions counterbalanced across participants to mitigate learning effects. The timeline of each test session is illustrated in Figure~\ref{fig:test-session-timeline}. During SAGAT pauses, the interface was hidden to prevent participants from using it to answer the questions.

At the end of each test session, participants completed the Bedford Workload and SART questionnaires. Finally, the TAM questionnaire was presented, and a semi-structured interview was conducted to gather qualitative feedback on the user experience, challenges faced, and suggestions for improvement.

\subsection{Results}
This section presents the results of the user study. We first report quantitative findings across situational awareness, performance, and workload, followed by qualitative insights into how pilots used the interface, adapted their strategies, and identified opportunities for scalability.
\subsubsection{Quantitative Results}

The quantitative results are reported in the order of 5$\rightarrow$10$\rightarrow$15 drones.

\begin{figure}[t]
\centering
\Description{A grid of five scatter plots with dashed trend lines showing quantitative study results across fleet sizes of 5, 10, and 15 drones (x-axis). Panel A (Mission SAGAT Score): three trend lines for Perception (Per, blue circles), Comprehension (Comp, orange squares), and Projection (Proj, green triangles); Perception increases from 0.65 to 0.80 with fleet size while Comprehension and Projection decline. Panel B (Drone SAGAT Score): same SA-level breakdown; all three levels drop from 5 to 10 drones (from ~0.85–0.88 to ~0.56–0.75), with slight recovery at 15. Panel C (Cognitive Load): Bedford Workload Scale mean scores increase monotonically from ~4 to ~6.3. Panel D (SART): mean subjective situational awareness scores decrease monotonically from ~7.7 to ~4.8. Panel E (Per-Drone Productivity): mean inspection waypoints covered per drone decrease from ~39.6 (5 drones) to ~38.2 (10 drones) and ~32.1 (15 drones).}
\includegraphics[width=\columnwidth]{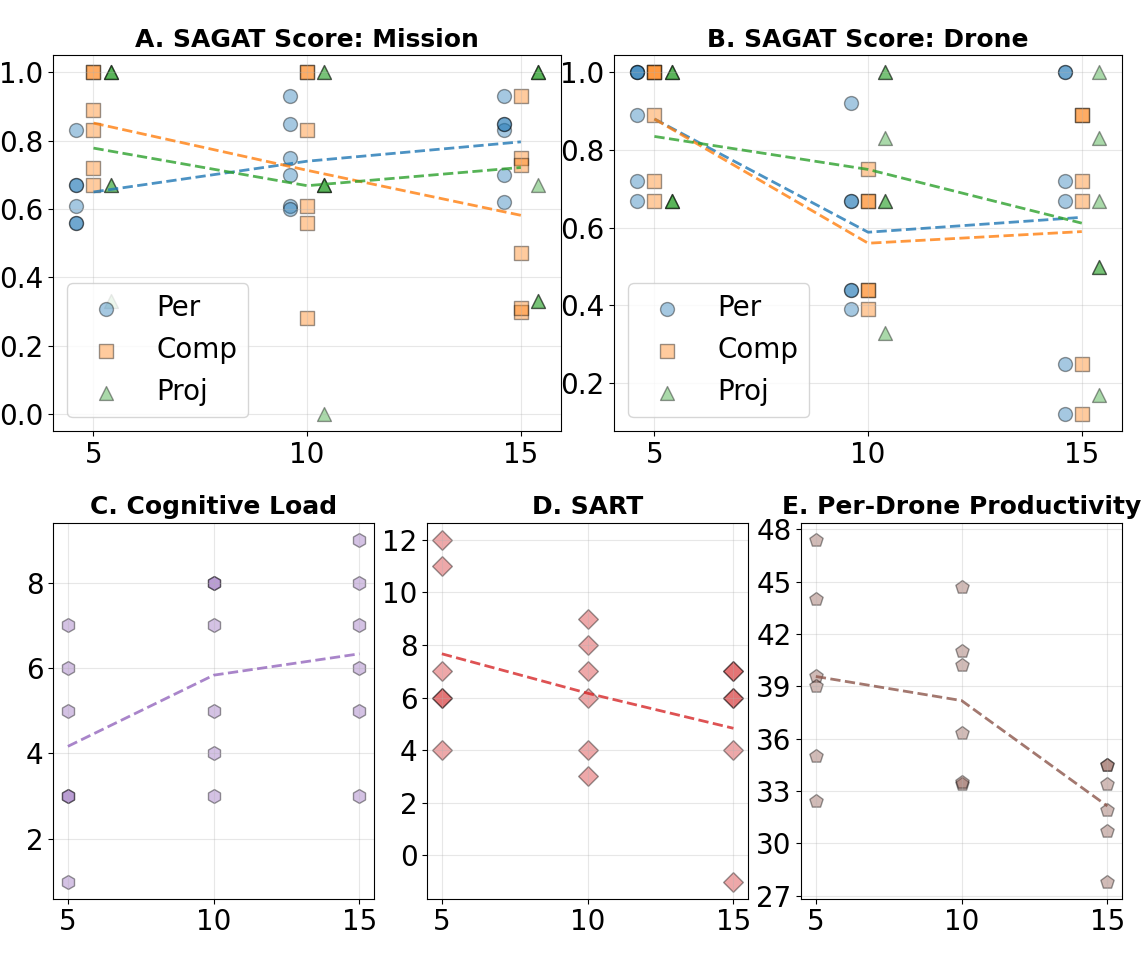}
\caption{SAGAT scores broken down by SA levels (Perception, Comprehension, Projection) for the (A) Mission Progress and (B) Drone Status domains; Performance logging data of (C) Per Drone Productivity; Subjective questionnaires: (D) Bedford Workload Scale, and (E) SART.}
\label{fig:quantitative-data}
\end{figure}

\paragraph{Mission progress situational awareness.}
As shown in Figure~\ref{fig:quantitative-data}(A), mission-level situational awareness presents mixed trends across SA levels as fleet size increases. 
Mission perception (level 1) increases from 0.65 (SD=0.10) to 0.74 (SD=0.13) and 0.80 (SD=0.12).
In contrast, mission comprehension (level 2) decreases from 0.85 (SD=0.14) to 0.71 (SD=0.28) and 0.58 (SD=0.26).
Mission projection (level 3) decreases from 0.78 (SD=0.27) to 0.67 (SD=0.37), then partially recovers to 0.72 (SD=0.33).
Overall, average mission SA across the three levels decreases from 0.76 (SD=0.12) to 0.71 (SD=0.21) and 0.70 (SD=0.19).

\paragraph{Drone and safety situational awareness.}
In the drone status domain (Figure~\ref{fig:quantitative-data}.B), all three SA levels declined with fleet size compared with the 5-drone condition. Drone perception dropped from 0.88 (SD=0.15) to 0.59 (SD=0.20), with a slight recovery to 0.63 (SD=0.37).
Drone comprehension followed a similar pattern, decreasing from 0.88 (SD=0.15) to 0.56 (SD=0.15) and 0.59 (SD=0.33). 
Drone projection decreased from 0.83 (SD=0.18) to 0.75 (SD=0.25), then further to 0.61 (SD=0.29). 
Aggregating the three SA levels, the overall drone/safety SA declined from 0.86 (SD=0.11) to 0.63 (SD=0.17) and 0.61 (SD=0.24). 

\paragraph{Per-drone performance.}
Per-drone inspection throughput (Figure~\ref{fig:quantitative-data}.E) indicates a small decrease from 5 to 10 drones, followed by a larger decrease from 10 to 15 drones. The mean covered waypoints per drone were 39.6 (SD=5.54), 38.2 (SD=4.53), and 32.1 (SD=2.60). 

\paragraph{Subjective situational awareness and cognitive load.}
Subjective ratings show a clear inverse pattern between perceived SA (Figure~\ref{fig:quantitative-data}.E) and workload (Figure~\ref{fig:quantitative-data}.D). Mean SART scores decreased monotonically from 7.67 (SD=3.14) to 6.1 (SD=2.32) and 4.83 (SD=3.06). 
In parallel, Bedford cognitive load increased from 4.17 (SD=2.23) to 5.83 (SD=2.14) and 6.33 (SD=2.16)
, indicating progressively higher mental demand and lower situational awareness.

\begin{figure}[t]
\centering
\Description{A diverging horizontal stacked bar chart showing Technology Acceptance Model (TAM) questionnaire responses from 6 participants on a 7-point Likert scale. Four constructs are shown from top to bottom: PU (Perceived Usefulness), PEOU (Perceived Ease of Use), ATT (Attitude Toward Using), and BI (Behavioral Intention). Bars extend left for lower ratings (1–3, shades of dark red and orange) and right for higher ratings (5–7, shades of light and dark blue); neutral rating 4 is shown in gray. Most responses for all four constructs are concentrated in ratings 5–7, indicating overall positive acceptance. PU has 1 negative response (at rating 2), while BI shows the most uniformly positive distribution.}
\includegraphics[width=\columnwidth]{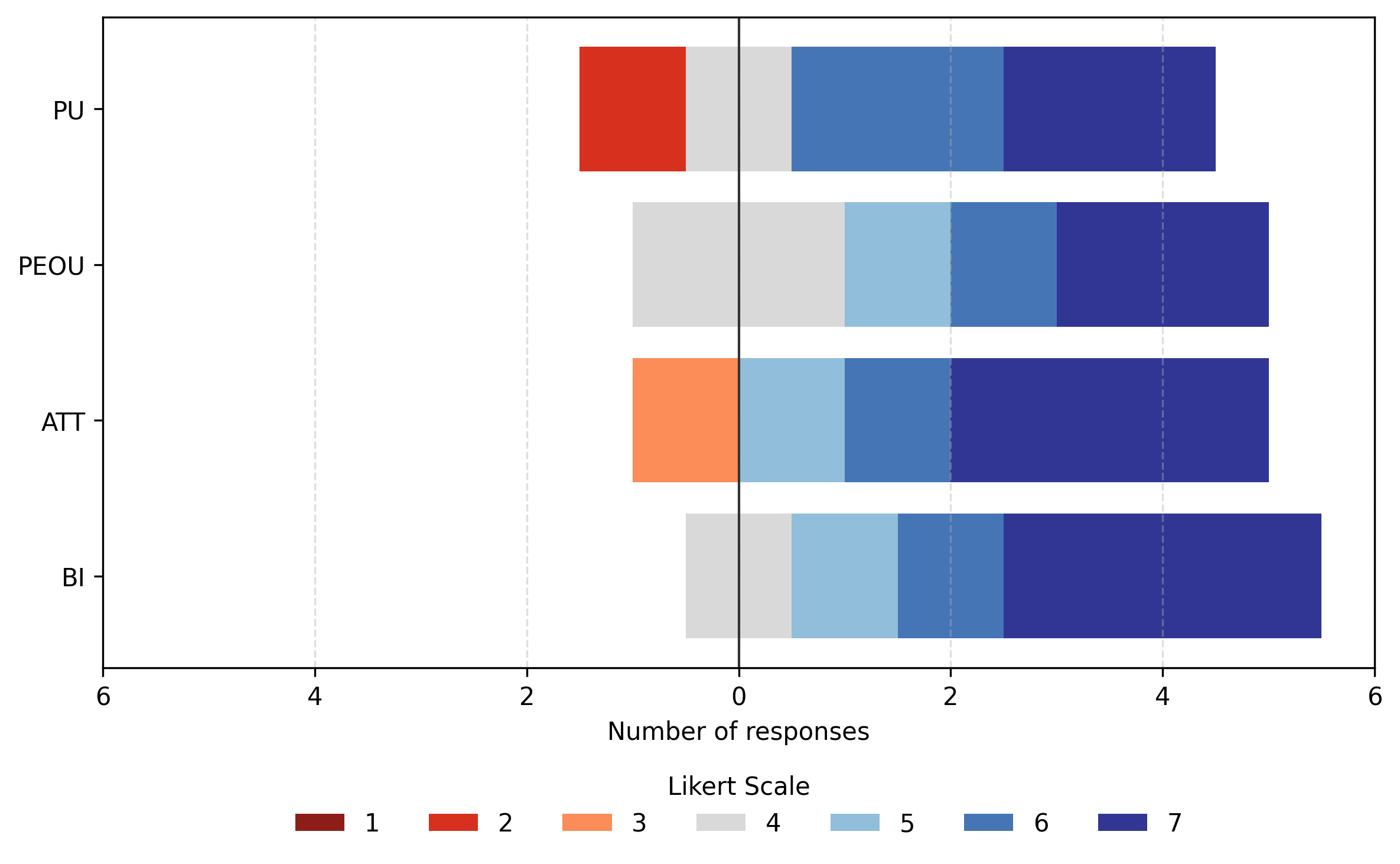}
\caption{Technology Acceptance Model (TAM) questionnaire results. PU denotes perceived usefulness, PEOU denotes perceived ease of use, ATT denotes attitude toward using, and BI denotes behavioral intention.}
\label{fig:tam}
\end{figure}

\paragraph{Technology Acceptance} The results (Figure~\ref{fig:tam}) show a generally positive reception of the interface, with a mean score of 5.33/7 for PU (SD=1.97), 5.5/7 for PEOU (SD=1.38), 5.83/7 for ATT (SD=1.60), and 6/7 for BI (SD=1.26). 

%

\subsubsection{Qualitative Results}

We analyzed participants' feedback from both the trial stage (labeled as "t") and the final stage (labeled as "f"). We present the results organized into five themes:

\paragraph{Interface usage patterns.} Participants consistently reported using the Sandtable as their primary interface. The Fleet Status Board was mainly used for responding to alerts, while the Minimap served as a second overview, as detailed by f2\_l: \textit{"I find myself using the 3D view the most, often for completion progress. Then status board is just to monitor the status of each drone."}
 
 Pilots described the Minimap as being useful for quick orientation but less informative than the Sandtable due to the lack of depth cues (t3\_l). Similarly, f1\_l reported relying less on the Minimap during active control, but using it when spatial alignment was unclear in 3D: \textit{"sometimes the 3D map I can't see where I need to rotate (align). At that situation I will look at the 2D map"}.

\begin{figure}[t]
\centering
\begin{subfigure}{0.602\columnwidth}
  \centering
  \Description{Screenshot of the FleetScape MR interface showing participant f2\_l's panel layout: Minimap displaying an overhead city map occupies the left side; Sandtable in the center; the vertical Fleet Status Board occupies the far right; a small camera view window floats above the center of the Sandtable. This arrangement places all four panels in a wide arc around the operator.}
  \includegraphics[width=\linewidth]{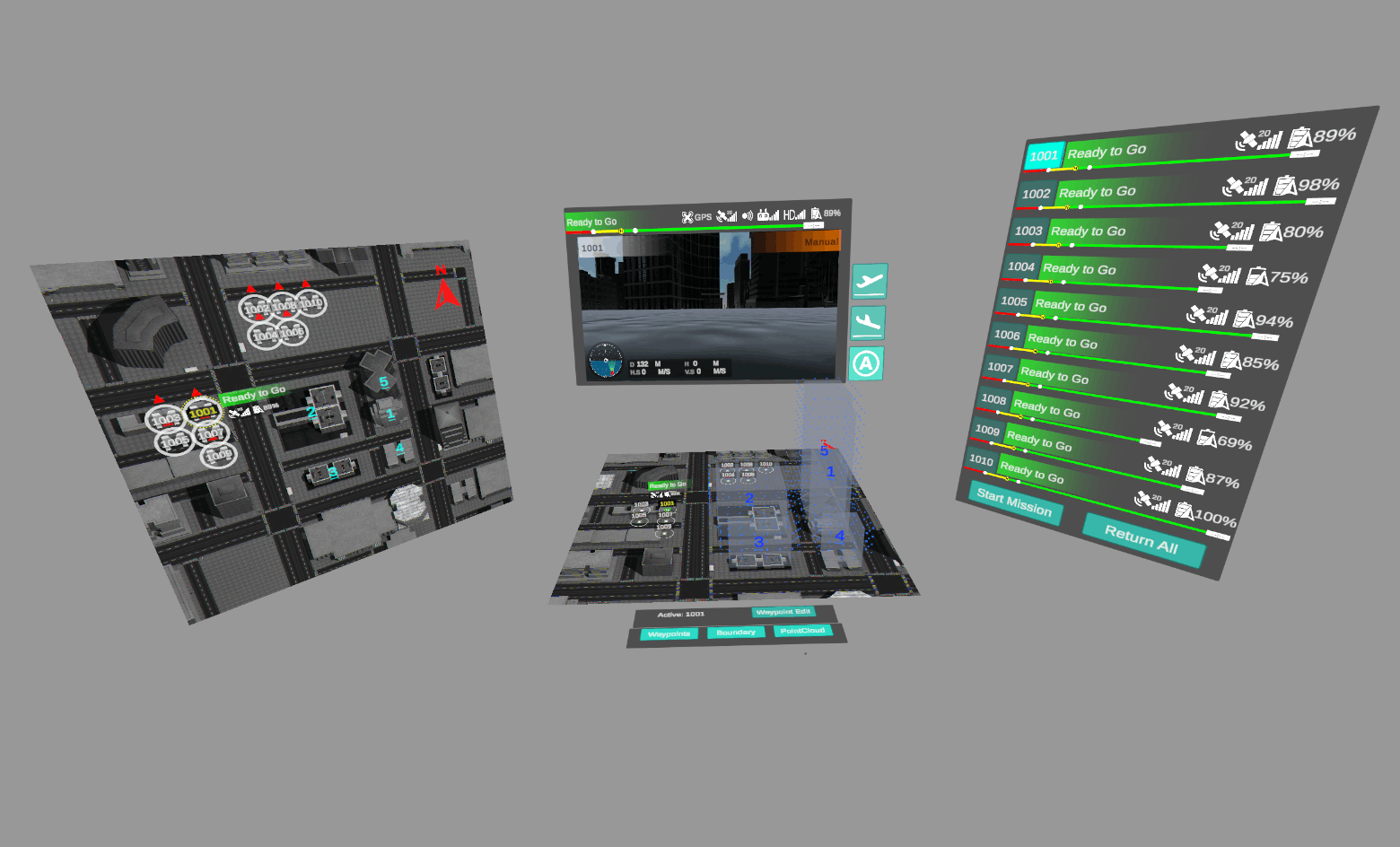}
  \caption{Final Layout of f2\_l.}
  \label{fig:layout-f2}
\end{subfigure}
\hfill
\begin{subfigure}{0.388\columnwidth}
  \centering
  \Description{Screenshot of the FleetScape MR interface showing participant f5\_l's panel layout: the Fleet Status Board is positioned at the upper right; the Sandtable occupies the center foreground with a different rotational orientation; the camera view is placed to the left of the Sandtable; and the Minimap is positioned overlapping the lower-left region of the Sandtable. The overall arrangement is more compact and vertically stacked compared to the typical layout.}
  \includegraphics[width=\linewidth]{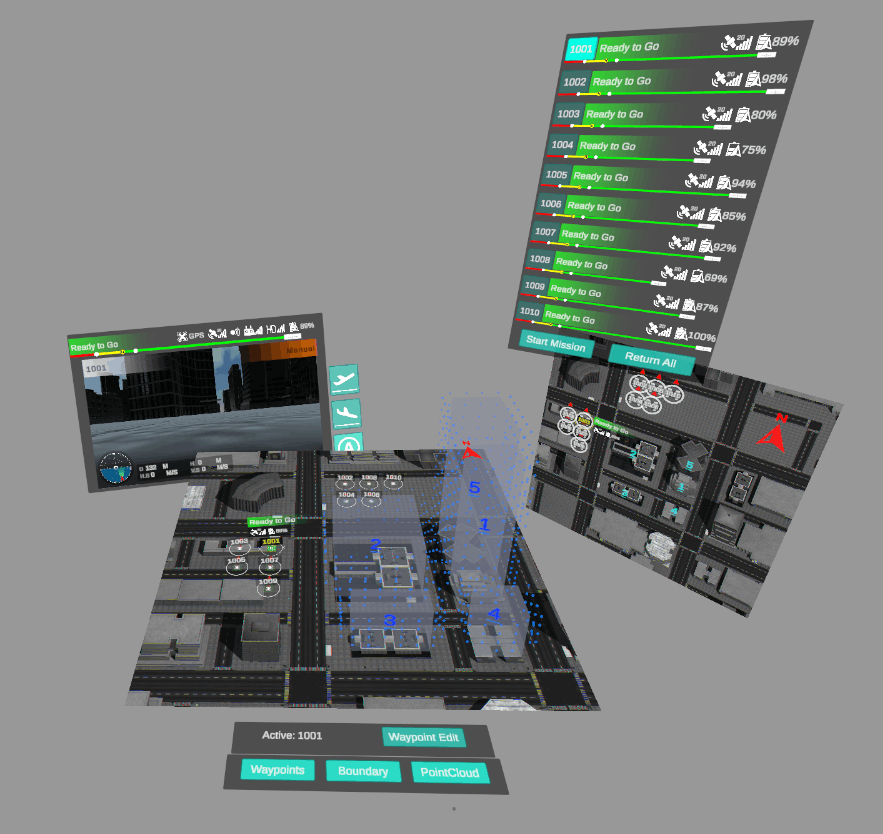}
  \caption{Final Layout of f5\_l.}
  \label{fig:layout-f5}
\end{subfigure}
\caption{Examples of spatial layout. (a) A typical layout most pilots used exemplified by f2\_l. (b) A specific layout based on personal preference (f5\_l). }
\label{fig:layout}
\end{figure}

 As for the spatial layout (shown in Figure~\ref{fig:layout}), most pilots placed the panels into a preferred position before the mission started. A layout of Minimap on the left, Sandtable in front, Fleet Status Board on the right, and the camera view on top of the Sandtable was adopted by most pilots. Two pilots frequently adjusted the layout during the mission (f5\_l, f6\_a). F5\_l detailed that: \textit{"I try to look at both (the sandtable and the camera view). 
 ... I keep adjusting the screens... because I want everything to be within my eyesight."} 

\paragraph{Situational awareness through the Sandtable.} 

Participants mostly enabled all Sandtable visualization: Inspection Waypoints, Point Clouds, and Building Boundaries, except for one pilot who switched off the Point Clouds to reduce visual cluttering (f2\_l). All visualizations were considered useful for different purposes. Inspection Waypoints and trajectories helped participants understand the drones' destinations and remaining coverage (f1\_l, f2\_l, f5\_l), while Point Clouds were used to infer scanned areas as well as scan coverage and distribution (f1\_l, f3\_l, f6\_a). 
Some participants found the Point Clouds distracting or unnecessary (f2\_l), and in some cases confusing with Inspection Waypoints, leading to uncertainty about drone behavior (f4\_l). Building Boundaries supported the understanding of mission regions and collision avoidance. For instance, f1\_l mentioned that the Building Boundaries and the Inspection Waypoints helped gauge the safety distance when placing custom waypoints, and f2\_l found following the Building Boundaries for collision avoidance was safer than the point cloud. 

The Sandtable also enhanced awareness of drone status and surroundings. F1\_l stated that: \textit{"I can see better [on the Sandtable] with the position of the drone… I can move around and look at it… where is it, part of the building… [so] I have more confidence in just flying the drone without looking at the live screen."}. 
F5\_l frequently toggled drone status on the Sandtable, reducing attention switches with the Fleet Status Board, and f6\_a reported that it helped reduce crowding and collision risk among drones. 


\paragraph{Interaction across control levels.} Participants used spatial interactions on the Sandtable, including selecting drones, zooming in/out, rotating, and translating, to shift between fleet-level overview and focused attention on one specific drone. Participants appreciated zooming the Sandtable to a specific drone or building (t1\_a, f5\_l, f6\_a). 

During manual control, several pilots mentioned that the Sandtable was useful for planning and the camera view for executing movement in space (t2\_a, f2\_l). F2\_l reported: \textit{"When I manually control the drone, I use the camera view to navigate, but I also use the 3d view to see where I want it to be.\"}. Other participants noted that the same 3D space can be used to place and adjust waypoints for rerouting the drone semi-autonomously. Some reported that the Building Boundary and the Inspection Waypoints provided spatial constraints and guidance (t2\_a, f1\_l).

Several pilots highlighted an interaction loop of autopilot supervision, alert handling, and re-routing. F2\_l described an alert-based strategy: \textit{“I will only take manual control when the autopilot can't solve the issue… [e.g.,] GPS position loss: I will manually fly the drone to a more open space… resume the auto mission… [or] manually overwrite the waypoint to a smaller building… Other than that, I would just leave it to auto”}, indicating a shift toward reactive behavior. F1\_l reinforced the advantage of spatial interaction in this interaction loop and stated that: \textit{“Because I know when I click on the status spot [for] the drone number, the screen will show the feed… From there, I can do manual control, and I will rely actually greatly on the sand table… [for] spatial control, like in between two building[s].”}. 

\paragraph{Adapting to different fleet sizes and supervision strategies.}

Participants differed on the optimal fleet size due to varying reasoning and perceived trade-offs. Most professional pilots considered 5 drones to be realistic for real-world flight operations. F3\_l stated that: \textit{"In real world, I tend to not push the boundary ... I will say five will be comfortable, 10 will be pushing it, 15 I only do in a simulator."} Others favored 10 drones for balancing the efficiency of the inspection and the cognitive load (f1\_l, f5\_l), while some amateurs preferred 15 drones to maximize coverage (t2\_a, f6\_a). F6\_a detailed this by stating: \textit{"For 15 drones I would prefer it because that gives you the most satisfaction that covered most of the spaces."} 

Fleet size also influenced control strategies, with more manual intervention observed for 5 drones and greater reliance on autonomy for 15 drones. F2\_l explained that: \textit{“When there are less drones… I can prioritize the strategy… change the waypoint manually and let it finish the smaller buildings first… When the fleet size is too large, my priority is to keep everyone safe… I don't have a lot of resources to optimize the strategy.”}
Instead, f6\_a used a different approach by pre-placing custom waypoints to avoid chaos in 15-drone scenarios. He explained that: \textit{"I actively put the waypoints before the drone started flying. By default, if you don't do anything, I think it'll create a lot more trouble."}


For 15 drones, participants reported a loss of detailed awareness due to high attentional demands and limited time (t1\_a, f1\_l, f3\_l). F1\_l detailed that: \textit{“For 15 drone[s]… attention needed for the whole operation is too much… So for 15 I think it's not [good] to use the manual piloting of the drone… just let it go; it will be easier that way.”}
F3\_l described a mental checklist loop to maintain situational awareness: \textit{“When I get more and more drones, I need to switch between what I want to check. So basically, making my drone is well spread, paying attention to how many surfaces are finishing and the buildings, listening to the audio cue for collision or position loss, and checking the batteries. So I have to keep switching, like, mentally, going through a checklist: 1‑2‑3, 1‑2‑3.”}. 

\paragraph{Design opportunities for scaling to larger fleets}

Participants suggested several opportunities for redesign to better support large fleets. T3\_l suggested focusing on reporting only complex errors to solve, to reduce the operator burden. Both f3\_l and f4\_l emphasized the need for faster interaction, possibly leveraging hand gestures or an additional tangible touch screen to manipulate the map.

Several pilots suggested that the system should support pre-mission planning (f2\_l, f3\_l, f4\_l). In particular, f2\_l suggested planning the drone paths before taking off. F4\_l advocated detailed scheduling of take-off timing, and dividing buildings into smaller regions to facilitate manual drone assignment: \textit{"If I would know where it is going in the first place and which surface will start first, then I will be more aware of where my attention (should go)."} F4\_l and f5\_l both suggested creating control groups for drones. For example, clustering drones by the building they are inspecting, so that the pilot can monitor and control at the group level instead of individual drones. F5\_l also suggested using colors to code the drone groups and buildings to support situational awareness.

\section{Discussion}

Our results suggest that multi-drone supervision can be understood as a form of \textit{spatial interaction} structured around three interdependent challenges: maintaining situational awareness, enabling fluid transitions across control levels, and supporting scalability.
Across both quantitative and qualitative findings, participants reported a strong overall acceptance of the system. TAM results indicate high perceived ease of use, positive attitudes towards spatial interaction, and a firm intention to adopt the interface, despite one instance of skepticism regarding perceived usefulness due to real-world workflow differences. Participants consistently emphasized the value of grounding fleet supervision in a shared 3D spatial representation, particularly for understanding spatial relationships and maintaining situational awareness across control modes. 

At the same time, both performance and situational awareness measures revealed clear scaling effects. While most pilots preferred operating 5-10 drones, they were able to manage a fleet of 15 drones by adapting their strategy to an exception-based handling approach.
Building on our findings, we derive and discuss four design implications for scalable spatial interfaces.


\paragraph{Spatial representation for safety and mission awareness.} 

Our results show that participants relied heavily on the Sandtable as their primary source of information, with relatively stable mission-level situational awareness despite increasing fleet size, and a reduction in drone-level (safety) awareness. This suggests that spatial representations effectively support global understanding but become more challenging for tracking individual elements at scale.


The Sandtable integrates heterogeneous data layers into a shared spatial frame. Participants' feedback suggests that this reduces the need for mental rotations~\cite{shepard1971mental} and enables direct perception of altitude, geometry, and proximity. Qualitative feedback further indicates that different visualization layers support different aspects of situational awareness, which is consistent with the SAGAT results. For Level 1 mission SA (perception), participants found the real-time point cloud highly effective for instantly tracking mission progress and identifying active work areas. This effectiveness is reflected in our quantitative results, where mission perception scores increased as fleet size grew. For spatial reasoning and manual navigation, operators heavily relied on the building boundaries to gauge safety thresholds and physical constraints, which helped Level 2 safety SA (comprehension). Meanwhile, the combination of inspection waypoints and planned trajectories actively supported Level 3 mission and safety SA (projection) by allowing pilots to intuitively anticipate where drones were heading next and to plan potential reroutes. 



Participants selectively relied on different layers depending on their task, and exhibited individual preferences in how these layers were combined. This suggests that spatial representations should not be treated as static views, but rather as configurable combinations of task-dependent layers. Interfaces should therefore support rapid filtering and adaptation of spatial information to align with operator goals and strategies.


\paragraph{Spatial interaction for fluid transition between manual intervention and fleet supervision.}

Qualitative observations revealed a recurring interaction loop across autonomy levels: operators monitor the fleet, detect an issue, zoom to focus on a specific drone, intervene via manual control or waypoint editing, and then return to global supervision. The ability to perform this loop within a shared spatial context was central to the interaction design. Participants leveraged spatial manipulation (zooming, rotating, translating) to move between levels of attention, enabling rapid localization of issues and coordination of actions. 
These observations suggest that spatial interaction supports both navigation and the integration of monitoring and intervention.



However, our results also highlight limitations in interaction efficiency. Participants reported difficulties in selecting moving drones in dense environments, which required them to switch to the status board, thus disrupting their spatial workflow. Participants also reported that the controller-based input sometimes created challenges for precise manipulation, mostly in situations requiring speed over precision. These issues were more pronounced as fleet size increased.
This suggests that the effectiveness of spatial interaction depends not only on representation but also on the efficiency and robustness of interaction techniques. Future systems should explore alternative input modalities (e.g., tangible or gesture-based interaction) and adaptive control mechanisms to better support fast and precise operations at scale.

\paragraph{Adaptive visualization to balance detail and clutter for larger fleets.}

Our quantitative results show a divergence between situational awareness levels as fleet size increases: mission-level perception improves, while comprehension and drone-level awareness degrade. Our findings suggest that pilots' capacity for monitoring the drone status had reached saturation between 5 and 10 drones, causing degradation to safety situational awareness for events such as low battery, GPS loss, or collision risk. Qualitative feedback further confirmed that interpreting detailed information about individual drones and local states was made difficult by dense visuals. 

These findings suggest a possible trade-off between visibility (level 1 SA) and interpretability (level 2 SA) in spatial interfaces as reflected in decreasing comprehension scores. Especially when Point Clouds and geometries are visually dense or ambiguous, participants fell back to the 2D overview for tracking states, leading to increased attention switches and cognitive efforts for extracting meaning, particularly in safety-critical scenarios.  


To mitigate this, future interfaces should incorporate adaptive visualization strategies that dynamically balance the level of detail based on context and scale. Adaptive strategies include providing aggregated or abstracted representations for larger fleets, while enabling pilots to drill down into finer details when needed. 
Furthermore, participants also suggested using grouping mechanisms and color-coding to differentiate between buildings and drones, instead of the numbers used. However, this may be challenging to implement as the number of drones increases~\cite{hoang2023}, in which case, grouping drones into control groups and color-coding the groups may be a more scalable solution. In addition, the visual channel can be augmented with multimodal feedback, such as audio alerts with voice-over, to provide more in-time critical information. 

\paragraph{Mission planning, abstraction, and prioritization to support scalability.}

Performance and qualitative results indicate a shift in operator strategy as fleet size increases. While participants actively optimized trajectories and coverage at smaller fleet sizes, they progressively shifted toward reactive, exception-based supervision at larger scales. This transition seems to match with the observed decrease in per-drone productivity and situational awareness as illustrated in Figure~\ref{fig:quantitative-data}.

This suggests that scalability cannot rely solely on improving real-time interaction, but requires restructuring the overall workflow. In particular, participants mentioned that greater emphasis should be placed on pre-mission planning, including route projection, task allocation, and identification of potential risk zones. By increasing predictability, such support can reduce the need for continuous intervention during execution.


In addition, abstraction mechanisms can help pilots manage complexity. Participants proposed grouping drones based on spatial or task-based criteria, allowing control and monitoring at a higher level. This aligns with prior findings in real-time strategy (RTS) games, where mechanics such as control groups can effectively help players manage large numbers of units~\cite{yan2015masters}. \added{The minimap design was also often featured in RTS games, making it easier for the player to hop around different areas of the map. However, adopting these design strategies in spatial interaction is often challenging due to the differences in input methods, users' perspectives, and the dimensionality of the environment~\cite{truman2018rethinking}. Moreover, drone fleet supervision and control blends higher levels of agent autonomy with higher degrees of control freedom, making it difficult to simply group the agents without accounting for the state of each individual. Future work can utilize RTS game design as a useful framing for more scalable fleet management.}


Finally, prioritization mechanisms are essential to guide operator attention toward the most critical events, particularly under high workload. For example, compound safety-critical events such as a low battery combined with GPS loss should be prioritized for alerts and visual emphasis, while other drones with less critical states can be deferred for later review. Future interfaces should therefore support selective highlighting and sorting of events based on urgency and impact.




\section{Limitations and Future Work}

Our work has several key limitations. First, the evaluation involved a relatively small sample size of expert drone pilots, limiting the statistical power of the quantitative results. Consequently, our findings are primarily exploratory, and the interaction patterns observed may not fully generalize to a larger population, including novice or moderately trained operators who might possess different cognitive load thresholds. 

Second, the study was conducted in a controlled simulated environment. Although this setup enabled a safe, isolated evaluation of fleet scaling, autonomy, and safety-critical events, which is common in the aviation industry, it inherently abstracted away critical real-world complexities. Physical deployments may introduce unpredictable elements such as \deleted{dynamic communication latency}\added{control and communication latency and reliability}, network bandwidth degradation, and environmental variability (e.g., wind gusts, varying lighting conditions). \added{The pilots may also have a reduced sense of jeopardy in the simulation, and their willingness to accept a technology is also confounded by factors such as airspace regulatory constraints, the reliability of the drone autonomy, and the maturity of the system infrastructure built to support it. The Fan-out estimate of 10.6 (Appendix \ref{sec:fleet-size-determination}) is based on interaction and activity times measured in simulation. In real-world deployment, latency, pose-error recovery, degraded feeds, and higher operational risk would likely increase interaction time and reduce autonomous activity time, thereby lowering Fan-out. We therefore describe 10.6 as an upper bound estimate under reliable autonomy conditions, and treat the observed 5–10 drones as the more conservative, practically meaningful range.} \deleted{In the real world, pilots' willingness to accept a technology is also confounded by factors such as airspace regulatory constraints, the reliability of the drone autonomy, and the maturity of the system infrastructure built to support it. }Still, our system design pushes one step forward towards real-world deployment of the multi-drone control system by providing practical insights on improving situational awareness and safety, and guidelines for scalability.

Finally, the experiment was constrained to a predefined 7-minute building inspection task\added{, which assumes careful planning, safety procedures, and routine, cooperative inspection goals, rather than emergency, military, or unknown environment scenarios}. It remains to be investigated how the interface and the spatial interaction techniques would perform during prolonged, continuous fleet supervision, or highly dynamic, time-critical tasks like search-and-rescue or active target tracking. 

\section{Conclusion}

Through the design and implementation of FleetScape, this paper provides evidence that drone fleet supervision can be framed as spatial interaction where telemetry, mission progress, and safety alerts are grounded within a unified 3D mixed reality sandtable, supporting situational awareness across multiple control levels. 
Our exploratory evaluation with expert pilots validated key aspects of our study. Participants benefited from the Sandtable for reducing mental rotation overhead and supporting rapid transitions between global supervision and local drone-level control. We identified an empirical limit of around 10 drones, beyond which visual clutter and cognitive load reduced situational awareness. Scaling beyond this limit requires adaptive visualizations, abstraction mechanisms such as control groups, and pre-mission planning to shift the cognitive demand from real-time intervention to workflow preparation. Our results provide actionable guidelines for designing effective and robust spatial interfaces for complex, real-world multi-robot operations.


\begin{acks}
This research is supported by the National Research Foundation, Prime Minister’s Office, Singapore under its Campus for Research Excellence and Technological Enterprise (CREATE) programme (DesCartes Project). We would like to thank all the participants who dedicated their time and efforts in our design workshop, pilot studies and final experiments and provided valuable insights. In particular, we thank Gautier, Murat, Nicholas, Lorelia for their suggestions in the early stages, and Wan Yi, Zheng Hao, Tingfang for their continuous feedback during design iterations. 
\end{acks}

\bibliographystyle{ACM-Reference-Format}
\bibliography{1_reference.bib,software.bib}

\appendix

\section{System Implementation}

\subsection{Drone Autonomy} \label{sec:autonomy}

\begin{figure}[t]
  \centering
  \includegraphics[width=0.95\columnwidth]{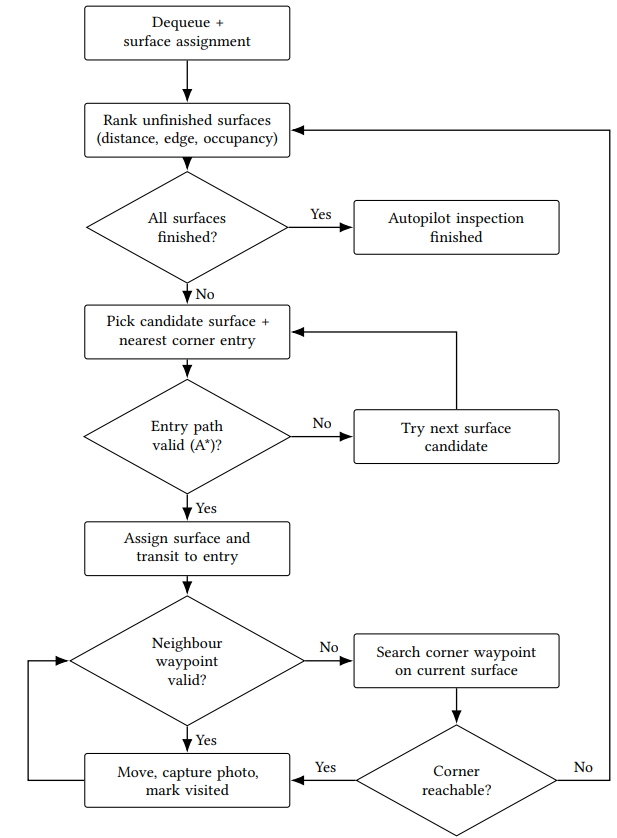}
  \Description{Flowchart of the per-drone autonomy loop. A drone waits for its turn in the autopilot queue and ranks unfinished surfaces. It then checks whether all surfaces are finished; if yes, the autopilot inspection ends. If not, it selects a candidate surface with a corner entry waypoint. If no A-star path exists, it evaluates the next candidate. If a path exists, it transits to the surface and performs neighbour-first waypoint inspection with photo capture. When no neighbouring waypoint is valid, it searches a corner waypoint; if reachable, inspection continues; otherwise, the loop returns to surface ranking to select a new candidate.}
  \caption{Autonomous inspection flow for each drone: queue-based task allocation followed by neighbor-first traversal with corner-waypoint fallback.}
  \label{fig:inspection_algorithm}
\end{figure}

Autonomy is implemented on a 2.5~m voxel grid with A* path planning. The drone agent uses A* search when transiting to surface entry points, indicated as "In Transit" on the status message. For task allocation, each drone ranks unfinished surfaces by distance to the surface plane and nearest edge, then selects the closest surface with the lowest occupancy. The drone will evaluate the cornered waypoint on the surface and select the closest as the entry waypoint. If no valid entry path exists, the drone evaluates the next candidate. To reduce inter-drone route conflicts, surface-assignment initialization is serialized through an autopilot queue. Once assigned, the drone follows a neighbor-first inspection routine. Photos are captured at each waypoint. When there is no valid neighboring waypoint, the drone will fallback to search for the corner waypoint of the surface, or hop to a new surface in the case of surface completion. Figure~\ref{fig:inspection_algorithm} summarizes the autonomous inspection loop executed by each drone.

\subsection{Navigation Uncertainty}

\added{
Uncertainty is quantized into an
integer level $\ell_i \in \{0, \dots, \ell_{\max}\}$ with $\ell_{\max} = 10$. The
reported scalar uncertainty is
\begin{equation}
  \rho_i(t) = \ell_i(t)\,\delta, \qquad \delta = 0.3\ \text{m},
  \label{eq:reported-uncertainty}
\end{equation}
so $\rho_i$ ranges from 0 to \SI{3.0}{\meter}. 
This is the value the interface renders around each sandtable miniature.
}

\added{
The level evolves as a biased random walk. About every \SI{0.5}{\second} (interval
drawn from a Gaussian with mean \SI{0.5}{\second}), $\ell_i$ is compared against a
regional target $\ell_{\text{region}}$ set by the uncertainty zone the drone
occupies. A random float $R$ is generated between $0.0$ and $1.0$. The level increments when $R$ is less than
\begin{equation}
  P(\ell_i \to \ell_i + 1) =
  \left(1 - \frac{\ell_i}{\ell_{\text{region}}}\right)
  \frac{\ell_{\text{region}}}{\ell_{\max}},
  \label{eq:level-walk}
\end{equation}
and otherwise, the level decrements when $R$ is less than $0.5$, with $\ell_i$ clamped to
$[0, \ell_{\max}]$. This models the gradual process of GPS loss and recovery in the real world. Instead of snapping to a target value, the pull toward $\ell_{\text{region}}$ weakens as the level
approaches it.
}

\added{
The backend reports a
sensed position to the frontend
\begin{equation}
  \tilde{\mathbf{p}}_i(t) = \mathbf{p}_i(t) + \mathbf{o}_i(t),
  \label{eq:sensed-position}
\end{equation}
where $\mathbf{p}_i(t)$ is the drone's true position in Unity, $\mathbf{o}_i(t)$ is a horizontal offset vector.
}

\added{
The offset $\mathbf{o}_i(t)$ tracks the current uncertainty. Its target magnitude
equals $\rho_i$ and its direction is random in the horizontal plane. The reported
offset slews toward this target at a fixed \SI{1.0}{\meter\per\second}, producing
smooth drift.
}

\added{
When $\rho_i$ exceeds an abnormal-signal threshold of \SI{1.0}{\meter}, the
interface marks the drone's signal as lost. An autonomous drone that is not already
returning to home then halts its autopilot with a ``Low GPS'' notice, requiring
operator intervention.
}

\section{Data Collection and Analysis} \label{sec:data-collection}

\subsection{Situation Awareness Rating Technique (SART)} \label{sec:sart}
For the subjective assessment of situational awareness, we utilized a short version of SART consisting of three questions, each rated on a 7-point Likert scale:
\begin{itemize}
    \item \textbf{Demand (D):} The task required a high level of attention and mental effort to keep track of the drones and the mission.
    \item \textbf{Supply (S):} I felt I had sufficient attentional resources to manage the drones and the information presented by the interface.
    \item \textbf{Understanding (U):} I had a clear understanding of what was happening in the system and how the mission was progressing.
\end{itemize}
The overall SART score is calculated using the formula: \text{SA} = $S + U - D$.

\subsection{Bedford Workload Scale} \label{sec:bedford-workload}
\begin{figure}[ht]
    \centering
    \includegraphics[width=\columnwidth]{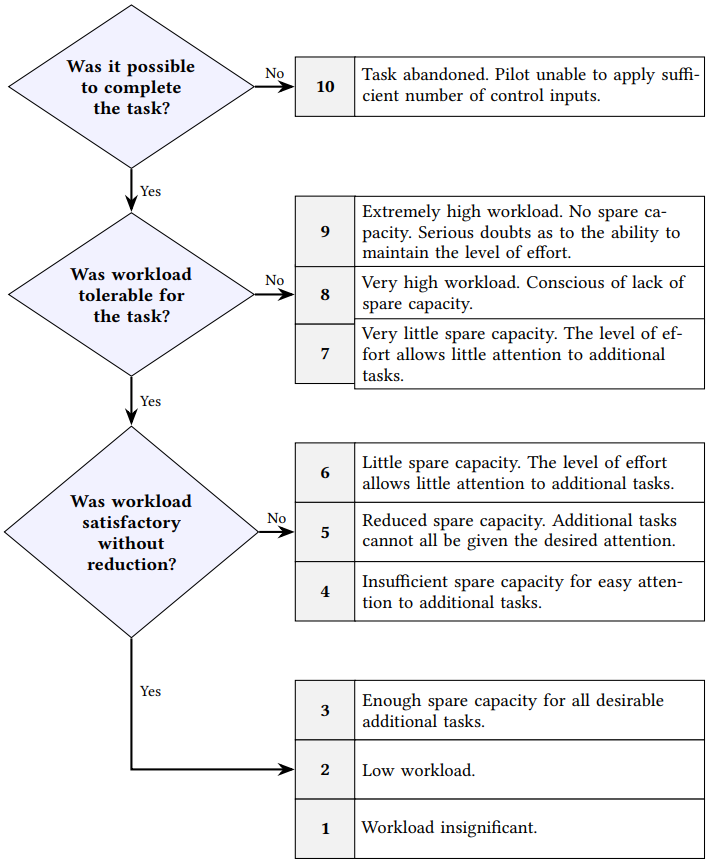}
    \Description{A hierarchical decision tree diagram for the Bedford Workload Scale. Three diamond-shaped decision nodes are arranged top to bottom. The first node asks, "Was it possible to complete the task?"; a No branch leads directly to rating box 10 (Task abandoned: pilot unable to apply sufficient number of control inputs). A Yes branch leads to the second node: "Was the workload tolerable for the task?"; a No branch spans ratings 7--9, describing very high to extremely high workload with no spare capacity. A Yes branch leads to the third node: "Was workload satisfactory without reduction?"; a No branch spans ratings 4--6, describing insufficient to little spare capacity. A Yes branch leads to ratings 1--3 describing low to insignificant workload. Each numbered rating box (1--10) is paired with a short text description of the corresponding cognitive workload level.}
    \caption{The full decision tree diagram for the Bedford Workload Scale.}
    \label{fig:bedford_scale}
\end{figure}

The Bedford Workload Scale uses a hierarchical decision tree (Figure~\ref{fig:bedford_scale}) to evaluate pilot workload on a uni-dimensional 10-point scale. Pilots answer a series of cascading yes/no questions regarding their spare mental capacity, which ultimately guides them to a specific workload rating.

\subsection{Technology Acceptance Model (TAM)} \label{sec:tam}
To evaluate users' acceptance of the interface, we used four questions based on TAM, scored on a 7-point Likert scale:
\begin{itemize}
    \item \textbf{PU (Perceived Usefulness):} The system is useful in my work.
    \item \textbf{PEOU (Perceived Ease of Use):} I find this interface easy to use.
    \item \textbf{ATT (Attitude Towards Using):} Overall, using this interface at work is a good idea.
    \item \textbf{BI (Behavioral Intention):} I intend to use this system in the future.
\end{itemize}

\subsection{SAGAT scoring} \label{sec:sagat-scoring}

Following Endsley's guidelines for direct SA measurement~\cite{endsley2017direct}, each SAGAT query is scored by comparing the participant's response against the ground truth state logged by the simulation at the moment of the pause. Specifically, for perception and comprehension questions where the ground truth is a set $G$ and the participant's response is a set $R$, the score is computed as:
\begin{equation}
\text{Score} = \frac{|G \cap R|}{|G \cup R|}
\end{equation}
This Jaccard index~\cite{niwattanakul2013using} penalizes both omissions (elements in $G$ but not in $R$) and false positives (elements in $R$ but not in $G$), yielding a score between 0 and 1.

For projection questions, the scoring is binary (0 or 1), based on the coherence between the answer and the reason given. 

SA scores are analyzed separately by level. We compute the average SA score for each SA level $l$ and domain $d$ by averaging the scores of the relevant queries:
\begin{equation}
\text{SA}_{l,d} = \frac{1}{N} \sum_{i=1}^{N} s_{i,l,d}
\end{equation}
where $N$ is the total number of observations and $s_{i,l,d}$ is the score for participant $i$ at SA level $l$ in domain $d$. 

\section{Research Methods}

\subsection{Fleet Size Determination} \label{sec:fleet-size-determination}
To determine appropriate fleet sizes for the study, we applied the Fan-out (FO) theory~\cite{olsen2004fan}. The Fan-out of a system is defined as:
\begin{equation}
\text{FO} = \frac{\text{AT}}{\text{IT}}
\end{equation}
where AT is the activity time (average interval between required interactions per drone), and IT is the interaction time (average time to complete an interaction, including situational awareness (SA) time).

\paragraph{Estimating IT.}
The three main interactions that pilots had with the drones were: (i) taking off idle drones after charging, (ii) manual control in response to collision warnings or GPS loss, and (iii) drones' trajectories re-planning for better inspection distribution. Based on timing data collected from three pilot studies, the average interaction times for these interactions were:
\begin{itemize}
    \item Take-off after charging (IT negligible): $\text{SA}_{\text{takeoff}} \approx 5$\,s
    \item Trajectory replanning: $\text{IT}_{\text{replan}} \approx 50.94$\,s
    \item Manual control: $\text{IT}_{\text{manual}} \approx 33.08 + 5\,\text{(SA)} = 38.08$\,s
\end{itemize}

\paragraph{Estimating AT.}
From our study design, the estimated event rates per drone were:
\begin{itemize}
    \item Take-off after charging: 1 per 526.39\,s
    \item GPS error: 0.4 per 7-minute session (420\,s)
    \item Replan: 0.4 per 420\,s
\end{itemize}

\paragraph{Computing FO.}
Since the three interaction categories have different interaction times, we compute a rate-weighted average IT. Denoting the rates as $r_{\text{takeoff}} = \frac{1}{526}$, $r_{\text{replan}} = \frac{0.4}{420}$, and $r_{\text{manual}} = \frac{0.4}{420}$\,s$^{-1}$:
\begin{equation}
\overline{\text{IT}} = \frac{r_{\text{takeoff}} \cdot \text{IT}_{\text{takeoff}} + r_{\text{replan}} \cdot \text{IT}_{\text{replan}} + r_{\text{manual}} \cdot \text{IT}_{\text{manual}}}{r_{\text{takeoff}} + r_{\text{replan}} + r_{\text{manual}}}
\approx 24.81\,\text{s}
\end{equation}
The combined interaction rate is $r = r_{\text{takeoff}} + r_{\text{replan}} + r_{\text{manual}} \approx 0.00380\,\text{s}^{-1}$, giving a combined activity time of $\text{AT} = 1/r \approx 263$\,s. Substituting into the Fan-out formula:
\begin{equation}
\text{FO} = \frac{\text{AT}}{\overline{\text{IT}}} = \frac{263}{24.81} \approx 10.60
\end{equation}
This suggests an optimal fleet size of approximately 10 drones. Based on this estimate, we selected three fleet size conditions---5, 10, and 15 drones---to evaluate pilot control strategies and performance across varying levels of task complexity.

\end{document}